# Composing Applications with OntoCompo

Christian Brel, Philippe Renevier-Gonin
I3S Lab (Université Nice-Sophia Antipolis / CNRS)
930 route des Colles, BP 145
06903 Sophia Antipolis Cedex, FRANCE
{brel, renevier}@polytech.unice.fr

## RESUME
Nous décrivons dans cet article un système utilisant des ontologies afin de composer des applications en préservant l'apparence des applications avant composition. Basé sur un processus de composition reposant sur la manipulation des Interfaces Homme-Machine (IHM) et utilisant des ontologies pour relier les tâches, les IHM et les fonctionnalités, l'outil, appelé OntoCompo, aide le développeur à composer les applications grâce à la sélection, l'extraction et le placement des différents éléments d'interface pour constituer la nouvelle application.

## ABSTRACT
In this paper, we present an ontology-based approach to compose applications while preserving their former look. Our composition process relies on the manipulation of User Interfaces (UI) and on several ontologies describing relationships between tasks, UI and Functionalities. Our tool, called OntoCompo, supports compositions realized by the developer thanks to the selection, extraction and positioning of UI elements to constitute the new application.

## 1. INTRODUCTION
The advent of web 2.0 and the apparition of a lot of "applications stores" introduce implicitly new needs for users and developers faced to this set of applications dis-posed on the web. Mash-up solutions [2] for example allow them to juxtapose several applications and use them together. They can have ideas for new functionalities creating a new application combining existing ones. Adapting applications to users' requirements may be done through composition of applications. Tools for composing former applications (and probably corresponding source code) should introduce developers' comfort and a reduction of the time-to-market for new applications by recycling former applications.

In this paper, we present our tool, OntoCompo, dedicated to easily realize new applications by composition of their User Interface. Our tool is based on a process in three steps (Selection/Extraction/Positioning) [3]. To be composed, the applications have to be separate in two parts: (i) the User Interface, visible and well-known part of the application and (ii) the functional core, underground part of the application. Due to this clear separation, the composition process lets the possibility to the developer to build the new application selecting, extracting and positioning UI part of former applications, one after another.

Our tool is based on the UI manipulations. From selected UI elements, our tool can generate recommendations throughout the composition process to back the user. At any time, the link between the UI elements and the functional part elements are preserved. To keep the consistency of application, the tool uses Task Models (TM) as links between UI and functional parts, leading to a three parts representation look like Model-View-Controller (MVC) pattern. This mapping between tasks, functionalities and UI elements are implemented as ontologies and recommendations for extending selections are based on semantic queries and rules.

In the first section, we describe related works about application composition, then, a brief case-study, and before to conclude, we describe our tool OntoCompo.

## 2. RELATED WORK
As we aim at composing applications by manipulating their UI, we have to decompose UI, i.e. describe UI in order to deal with subparts of former UI. The description of an UI both involves:

(1) The description of its structure, i.e. the listing of the different components used in the interface and the inclusion relationship, like UIML [1], ALIAS [9], UsiXML [7] or MARIA [10]

(2) The spatial positioning of these components. By analyzing the different layouts used in the UI toolkits, we identified three ways to position the components in an interface: the AbsoluteLayout with X and Y coordinates, the TableLayout to place a component in a grid and the RelativeLayout to express the positioning of two UI components relatively to each other.

To manipulate applications in order to compose them, there are currently three main approaches: (i) the composition could be triggered by the functional (i.e. business) part, (ii) the composition could be triggered by the users' goals (i.e. tasks to be performed by users) and (iii) the composition could be triggered by the UI.

Each trigger addresses a specific problem of composition: presentation and layout considerations at the UI level ([6]), behavior of the application at the functional level ([9]), user needs at the task level ([10]).

These works do not reuse complete architecture of the former applications. Either they compose and reuse UI as first concern without any consideration of the links between UI and Functional part either their first concerns are functionality or task and provide the new application by generate or re-generate the user interface.

Our originalities are (i) to consider links between UI, tasks and functionalities, (ii) to lead the developer by suggesting him and asking him about elements to keep for aiming at composition consistency and (iii) to reuse existing UI in order to preserve former developments, former designs and former practices.

# 3. CASE STUDY

We take the example of a Human Resource Manager in a firm with two available applications. The first one (on the bottom right part of Figure 3) is an application to retrieve social insurance information about an employee from her insurance number like her first name, last name, birthday, birthplace or family status, name and birthday of her children etc. The second application (on the top right part of Figure 3) is an intern application in the firm to retrieve general employee information from her last name and first name like her posts and assignments into the firm or her visit card. The problem for our manager is to retrieve information from both applications (for example for editing pay slip) without to have to swap between them (with a potential loose of information during the swapping). So, to answer to our manager's needs, we propose to compose the two applications selecting parts of former applications she wants to keep in order to obtain a functional application preserving former designs from existing applications.

# 4. A USER-CENTERED TOOL FOR APPLICATION COMPOSITION

The aim of OntoCompo is to give an easy way for the developer of application to reuse existing applications to constitute her new one. We consider that a composition driven by a checked selection is a guarantee to preserve the global consistency of the final application. So we choose to help the developer for broadening selection. In terms of context of development, the developer will be able to choose functionalities, tasks or UI layout as extension way for her selection and then for extraction.

## 4.1 Hypothesis on Former Applications

To reuse existing applications, our hypothesis is to let the developer doing composition through the interfaces of applications. In the UI research field, there is a strong recommendation of using a Task Model (TM) during requirements analysis. So, our approach is to express links between the description of application (both functionalities and UI) and the TM of the application to provide a better support to the developer during the composition by preserving theses links to aim a functional application at the end. We use semantic annotations (using OWL Light language [11]) for the description of applications. The first advantage is the possibility to apply rules on semantic annotations to deduce some information on the layout of UI from former applications to preserve the UI elements proximity during the composition. To keep such of information, we decide to work with RelativeLayout, well-known layout to express positioning between two UI elements. We are able to refine from RelativeLayout positions of a UI component, new RelativeLayout position. For example, from a left position and top position of an element towards a second one, we can deduce a top-left position. And we have another category of rules to deduce RelativeLayout positions from AbsoluteLayout positions or TableLayout positions.

The second advantage to use semantic annotations is the possibility to query these annotations with a specific semantic engine like CORESE [5] and SPARQL language [12] to obtain the different links between tasks, UI elements and functionalities in order to suggest the developer new selection guaranteeing the consistency of the final application.

To be able to reuse elements of the former application, we need a software organization authorizing selection, extraction and rejigging of such elements. We opt for applications based on component architecture like FRACTAL components [4].

Naturally, the applications can have a component assembly for their functional part but they have to use components for their UI too. Due to this component architecture, by browsing the component assembly and the UI component structure (window, containers and graphical components), we are able to deduce the links between functionalities and UI parts. Moreover, this choice leaves the possibility to recompose an assembly by disconnecting and reconnecting components. In fact, this would be useful to obtain a functional application to finalize the composition process. Consequently, for reusing former applications parts, we use component-based software development to manipulate functionality assemblies and component-based UI with Java Swing JComponent encapsulated in component (FRACTAL component in the implementation of our prototype) in order to manipulate real UI parts.

To conclude, applications to compose are expected to (Figure 1):

- Be written as component assembly.
- Have a clear separation between its UI and its functionalities.
- Have a definition of its Task Model.
- Be provided with semantic annotations description of links between Task Model, UI and functionalities.

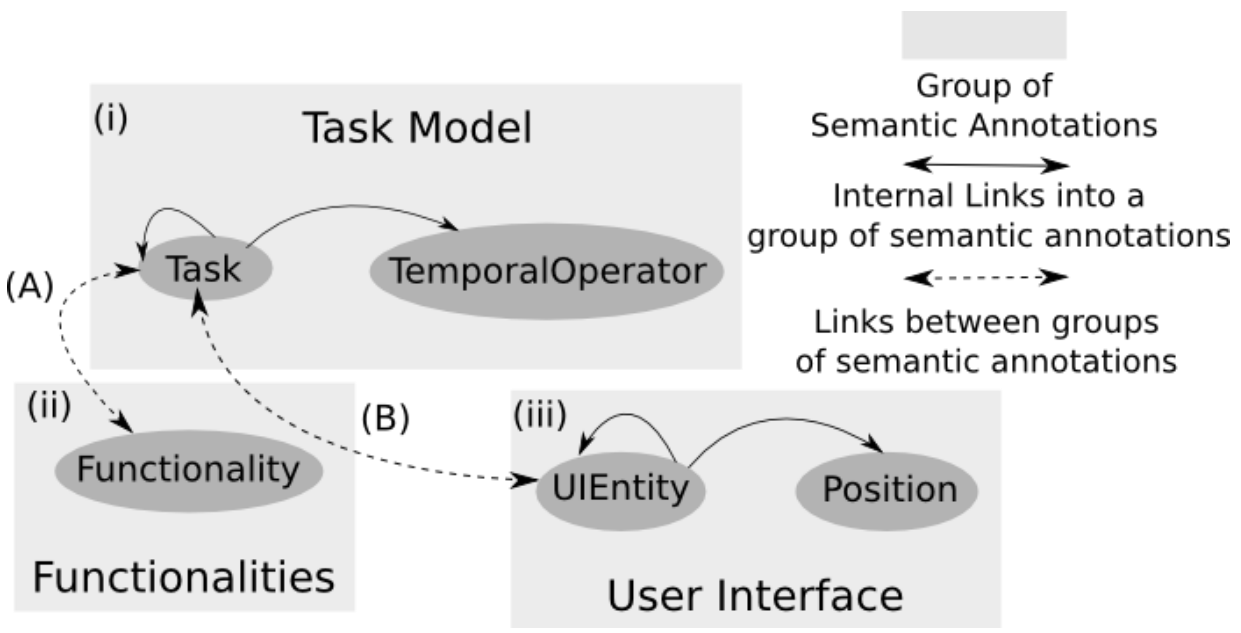

**Figure 1 Description of Applications loaded in OntoCompo**

## 4.2 Selection, Extraction and Positioning

After loading the applications, the entry point of our proposed process is the selection of the different UI elements on the UI. Selected UI elements are graphically highlighted. That simple selection is extended for performing complex selections or aiming at verifying consistency.

Figure 3 (left part) shows the main extra-interface for selection of UI elements the developer wants to keep for the new UI. We can find in the first part, annotated "S", different kinds of extensions the user can apply on current selection.

First, there is the layout extension. With the height toggle buttons for selected extension directions, the developer has the possibility to broaden the selection. That extension could be applied to the first selected component in current selection, to the last selected component in current selection or to all components in current selection. To perform this functionality, queries on semantic annotations are parameterized with the current selection (for example <BusinessDirSearchInputFC> indicated in Figure 3) and with each toggled direction (for example <OnTheRightOf>).

Secondly, there is the (container) parent extension. This extension uses queries on layout of application to obtain the parent container of last selected UI component in current selection. This extension allows the developer to be more efficient on her selection of all elements in a container potentially “hidden” by its contents.

Thirdly, there is the task extension. Each UI element is linked with a task described with semantic annotations. From the last selected component (for example <InsuranceCBirthDFC> in Figure 3), we use queries to obtain the task linked to it. From each returned tasks (here « Display Account Info » Task), we query semantic annotations to obtain all UI elements linked with this task. Retrieved UI elements are added to the selection (in our example all elements in <InsuranceCAccountInfoFC>).

Finally, there is the functionality extension. UI elements are directly linked to functionality but also through tasks. Since a task may be connected to several functionalities, it is possible to extend the selection to each part of the application by following these links. We start with selected UI elements. Thanks to SPARQL queries, we go back "up" to related tasks and then "up" to related functionalities. From these functionalities, we go back "down" to UI elements. Such retrieved UI elements are added to the current selection.

The developer can activate all theses extensions. She is free of combination between all proposed extensions. To help her and to lead her towards to a coherent composition, we develop a help selection. This help is a guide for the developer during all selection process. For each UI element, several questions suggest to the developer different possibilities for extending her selection. That help is controlled with the second part of the selection tools, annotated "H" in the Figure 3. The developer can use a help guided by tasks, by functionalities, by layouts or by a complete help (tasks, functionalities, layouts) to perform her selection. For this help, we use queries to retrieve the UI elements open to be added to selection.

When the developer is satisfied by her current selection, she has the possibility to extract it to an existing screen or a new screen (Figure 3, part "E"). For this extraction, for each UI element, we keep the links between tasks and functionalities in order to obtain a functional application and a reusable application for a possible future composition.

Finally, we provide a way to the developer to position UI elements for each screen. This positioning is based on RelativeLayout i.e. the elements can be visually position relatively to another one, by drag and drop. (Figure 2)

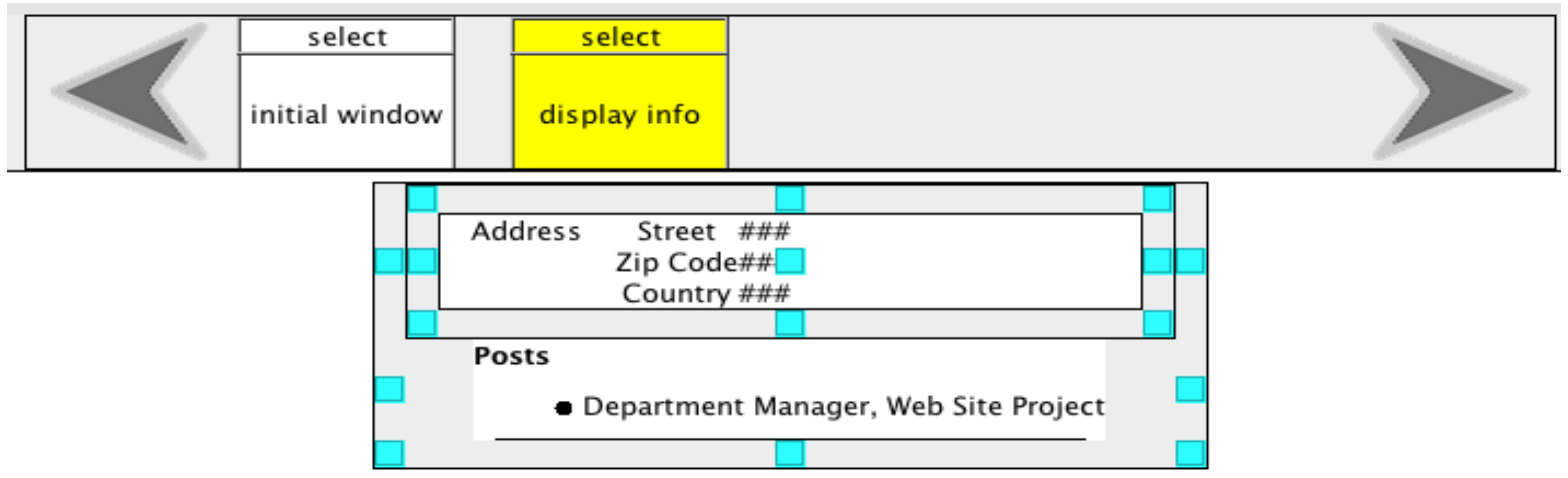

**Figure 2 Positioning Tool**

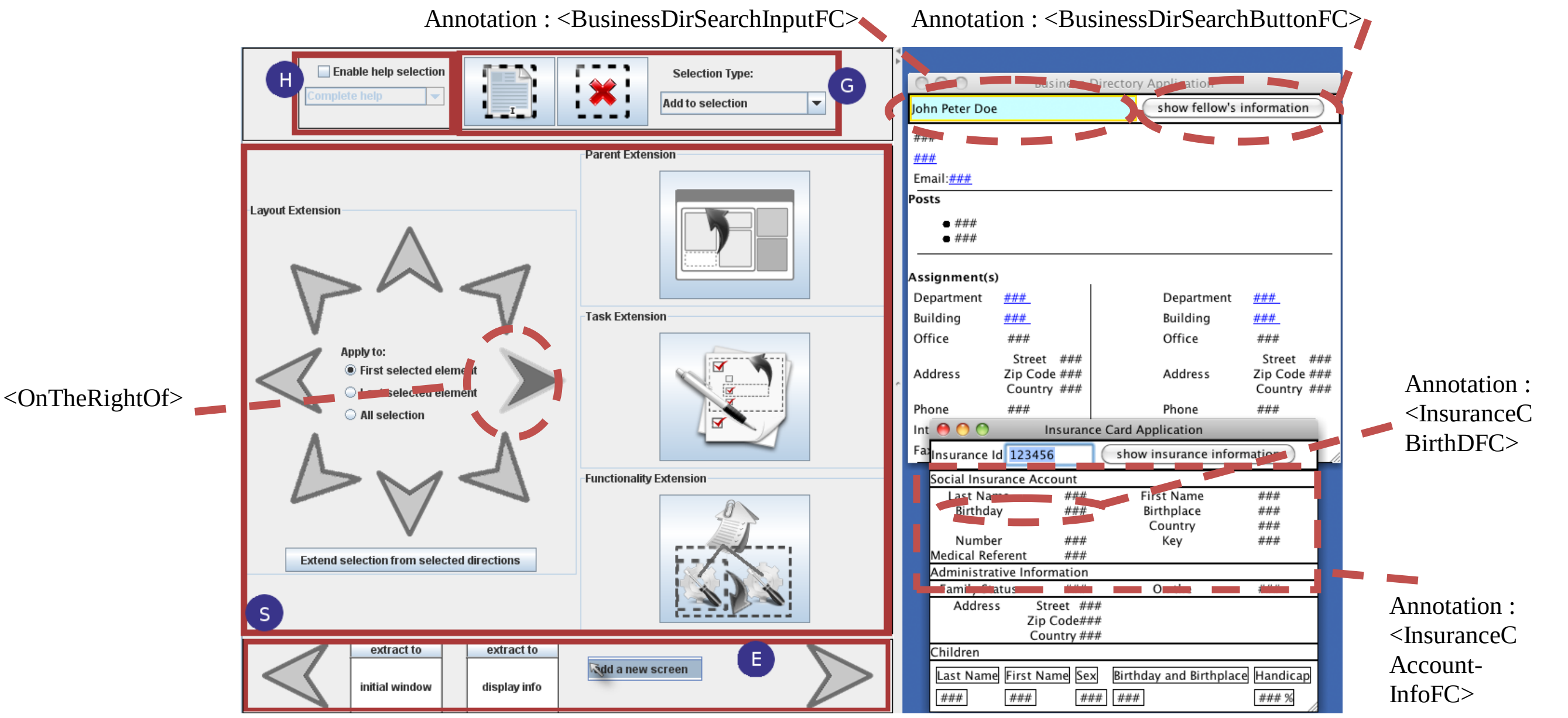

**Figure 3 <u>On the left part</u>: Tools for selection extensions and selection extraction -- <u>On the right part</u>: Case Study Applications' UI**

# 5. CONCLUSION

To conclude, in OntoCompo, we integrate help for the composition process based on these all selection extensions. Simple selection demonstrated a lack of efficiency and facility for the developer to build an aimed composition. So we propose a tool based on suggestions to extend the selection part of application to reuse. We took the decision to offer the developer a panel of extension by allowing him to choose her entry point (UI layout, functionalities or tasks) to perform her extensions. In this way, we are now planning user (developer) evaluation to validate the different entry point for this extension of selection. When they will be validated by user tests where we will observe the cognitive process of developers, we will be able to keep or give up the different extensions. Once that evaluation performed, we will work on a new step in the composition process about merging application elements (UI elements or functionalities).